\documentclass[11pt,a4paper]{article}
\pdfoutput=1
\usepackage{jheppub}
\usepackage{amsmath}
\usepackage{amssymb}
\usepackage{graphicx}
\usepackage{subcaption}
\usepackage{slashed}
\usepackage{cases}
\usepackage{tikz-feynman}
\usepackage[toc,page]{appendix}
\usepackage{empheq}
\usepackage[normalem]{ulem} 
\usepackage{xcolor} 

\allowdisplaybreaks

\preprint{}

\begin{document}

\title{How dark is the $\nu_R$-philic dark photon?}

\author[a]{Garv Chauhan}
\author[b]{, Xun-Jie Xu}

\affiliation[a]{Department of Physics and McDonnell Center for the Space Sciences,  Washington University, \\
St. Louis, MO 63130, USA}
\affiliation[b]{Service de Physique Th\'{e}orique, Universit\'{e} Libre de Bruxelles, Boulevard du Triomphe, CP225, 1050 Brussels, Belgium}

\emailAdd{garv.chauhan@wustl.edu}
\emailAdd{xunjie.xu@ulb.ac.be}

\date{\today}

\abstract{
We consider a generic dark photon that arises from a hidden $U(1)$
gauge symmetry imposed on right-handed neutrinos ($\nu_{R}$). Such
a $\nu_{R}$-philic dark photon is naturally dark due to the absence
of tree-level couplings to normal matter. However, loop-induced couplings
to charged leptons and quarks are inevitable, provided that $\nu_{R}$
mix with left-handed neutrinos via Dirac mass terms. We investigate
the loop-induced couplings  and
find that the $\nu_{R}$-philic dark photon is not inaccessibly dark, which could be of potential
importance to future dark photon searches at SHiP, FASER, Belle-II, LHC 14 TeV,
etc. 
}

\maketitle

\section{Introduction}

\noindent
Right-handed neutrinos ($\nu_{R}$), albeit not included in the Standard
Model (SM), are a highly motivated dark sector extension to accommodate
neutrino masses~\cite{Minkowski:1977sc,yanagida1979proceedings,GellMann:1980vs, glashow1979future,mohapatra1980neutrino},
dark matter~\cite{Asaka:2005pn,Asaka:2005an,Ma:2006km}, and baryon
asymmetry of the universe~\cite{Fukugita:1986hr}. Being intrinsically
dark, $\nu_{R}$ might have abundant new interactions well hidden
from experimental searches. In particular, it is tempting to consider
the possibility that there might be a hidden gauge symmetry in the
$\nu_{R}$ sector~\cite{Batell:2010bp,Chang:2011kv,Ma:2013yga,Lindner:2013awa,Ballett:2019pyw,Berbig:2020wve, Abdullahi:2020nyr,Jodlowski:2020vhr}. 
The new gauge boson arising from this symmetry does not directly couple to other fermions except for $\nu_{R}$ and
naturally becomes a dark photon, which we referred to as
the $\nu_{R}$-philic dark photon. 

The $\nu_{R}$-philic dark photon is not completely dark. It may interact
with normal matter via kinetic mixing \cite{Holdom:1985ag}, provided that the new gauge symmetry is Abelian; 
or, in the presence of mass terms connecting $\nu_{R}$ and left-handed
neutrinos $\nu_{L}$, via one-loop diagrams containing $W^{\pm}$/$Z$
and neutrinos~\cite{Pilaftsis:1991ug}.  In the former case, the strength of dark photon interactions
with quarks or charged leptons depends on the kinetic mixing parameter
$\epsilon$ in ${\cal L}\supset\frac{\epsilon}{2}F^{\mu\nu}F_{\mu\nu}'$
where $F^{\mu\nu}$ and $F_{\mu\nu}'$ are the  gauge field tensors
of the SM hypercharge $U(1)_{Y}$ and the new $U(1)$, respectively.
This case, being essentially independent of the neutrino sector, has
been widely considered in a plethora of dark photon studies---for
a review, see \cite{Essig:2013lka,Ilten:2018crw,Bauer:2018onh,Fabbrichesi:2020wbt}.
In the latter case, the loop-induced couplings depends on neutrino
masses and mixing, and will be investigated in this work.

The aim of this work is to address the question of how dark the $\nu_{R}$-philic
dark photon could be in the regime that dark-photon-matter interactions
dominantly arise from $\nu_{L}$-$\nu_{R}$ mixing instead of kinetic
mixing. We note here that the dominance might be merely due to accidentally
small $\epsilon$, or due to fundamental reasons such as the SM $U(1)_{Y}$
being part of a unified gauge symmetry {[}e.g.~$SU(5)${]} in grand
unified theories. We opt for a maximally model-independent framework
in which generic Dirac and Majorana mass terms are assumed. The loop-induced
couplings are UV finite as a consequence of the orthogonality between
SM gauge-neutrino couplings and the new ones.  Compared to our previous
study on loop-induced $\nu_{R}$-philic scalar interactions~\cite{Xu:2020qek},
we find that the couplings in the vector case are not suppressed by
light neutrino masses, and might be of potential importance to ongoing/upcoming
collider and beam dump searches for dark photons.

The paper is organized as follows: In Sec.~\ref{sec:fr}, we describe the relevant Lagrangian used in this work, 
reformulate neutrino interactions in the mass basis,
and discuss generalized matrix identities for UV divergence cancellation for later use. In Sec.~\ref{sec:loop_coupling}, we first derive model-independent expression for effective coupling of $Z'$ to charged leptons/quarks through one-loop diagram involving $Z$ and $W$ bosons, respectively. We then evaluate the coupling strength in three different examples. In Sec.~\ref{sec:mass}, we present a qualitative discussion about possible connections between the $U(1)_R$ gauge coupling and the mass of $Z'$. In Sec.~\ref{sec:pheno}, we present  constraints from a vast array of current and future experiments spanning from collider searches to astrophysical phenomena. We finally conclude in Sec.~\ref{sec:concl} with details of one-loop diagram calculations relegated to Appendix~\ref{appendix:A}.

\section{Framework}\label{sec:fr}

We consider a hidden $U(1)$ gauge symmetry, denoted by $U(1)_R$,
imposed on $n$ right-handed neutrinos. The gauge boson of $U(1)_{R}$ in this work is denoted by $Z'$. The relevant part of the Lagrangian for 
the $U(1)_R$ extension reads\footnote{Throughout the main text we exclusively use Weyl spinors for conceptual simplicity, while in
the Appendix we use Dirac/Majorana spinors for loop calculations.}: 
\begin{eqnarray}
{\cal L} & \supset & \nu_{R,j}^{\dagger}i\overline{\sigma}_{\mu}D_{j}^{\mu}\nu_{R,j}+\left[\frac{\left(M_{R}\right)_{ij}}{2}\thinspace\nu_{Ri}\nu_{R,j}+\left(m_{D}\right)_{\alpha j}\thinspace\nu_{L,\alpha}\nu_{R,j}+{\rm h.c.}\right]\nonumber \\
 &  & -\frac{1}{4}F'_{\mu\nu}F'^{\mu\nu}+\frac{1}{2}m_{Z'}^{2}Z'_{\mu}Z'^{\mu},\label{eq:lag1}
\end{eqnarray}
where $\overline{\sigma}\equiv(1,-\vec{\sigma})$ with $\vec{\sigma}$ being three Pauli matrices; $\alpha$ denotes flavor indices; $(i,\ j)=1,\ 2,\ 3,\ \cdots,\ n$; and
\begin{equation}
D_{j}^{\mu}=\partial^{\mu}-ig_{R}Q_{R,j}Z'^{\mu}\,.
\end{equation}
Here $g_{R}$ is the gauge coupling constant of $U(1)_{R}$ and $Q_{R,j}$
is the charge of $\nu_{R,j}$ under $U(1)_{R}$. Note that for most
general forms of $M_{R}$ and $m_{D}$, both the Majorana and Dirac
mass terms in Eq.~(\ref{eq:lag1}) break the $U(1)_{R}$ symmetry.
In addition, for arbitrary charge assignments of $\nu_{R,j}$ under
$U(1)_{R}$, the model would not be anomaly free. Nevertheless, one
can construct complete models in which $M_{R}$ and $m_{D}$ arise
from spontaneous symmetry breaking and the cancellation of anomalies can be obtained when several $\nu_{R,j}$'s have different charges 
with $\sum_{j}Q_{R,j}^{3}=0$---see the example in Sec.~\ref{sub:modelB}.
In this section we neglect these model-dependent details and focus
on the general framework proposed in Eq.~(\ref{eq:lag1}).

The Dirac and Majorana neutrino mass terms in Eq. (\ref{eq:lag1})
can be framed as 
\begin{equation}
{\cal L}_{\nu\thinspace{\rm mass}}=\frac{1}{2}(\nu_{L}^{T},\ \nu_{R}^{T})\left(\begin{array}{cc}
0_{3\times3} & m_{D}\\
m_{D}^{T} & M_{R}
\end{array}\right)\left(\begin{array}{c}
\nu_{L}\\
\nu_{R}
\end{array}\right),\label{eq:x-4}
\end{equation}
where $\nu_{L}=(\nu_{L,e},\nu_{L,\mu},\nu_{L,\tau})^{T}$ and $\nu_{R}=(\nu_{R,1},\nu_{R,2},\cdots)^{T}$
are column vectors. The entire mass matrix of $\nu_{L}$ and $\nu_{R}$
can be diagonalized by a unitary matrix $U$: 
\begin{equation}
\left(\begin{array}{c}
\nu_{L}\\
\nu_{R}
\end{array}\right)=U\left(\begin{array}{c}
\nu_{1,\,2,\,3}\\
\nu_{4,\,5,\,\cdots}
\end{array}\right),\ \ U^{T}\left(\begin{array}{cc}
0_{3\times3} & m_{D}\\
m_{D}^T & M_{R}
\end{array}\right)U=\left(\begin{array}{cc}
m_{1,\,2,\,3}\\
 & m_{4,\,5,\,\cdots}
\end{array}\right).\label{eq:U_trans}
\end{equation}
Here $\nu_{i}$ ($i=1,\ 2,\cdots,\ n+3$) denote neutrino mass eigenstates, with $m_{i}$ being
the corresponding masses. We refer to the basis after the
$U$ transformation as the \textit{chiral basis}, and the one before the transformation as the \textit{mass
basis}.

In order to facilitate loop calculations, we need to transform neutrino
interaction terms from the chiral basis to the mass basis. In the
chiral basis, we have the following neutrino interaction terms: 
\begin{equation}
{\cal L}\supset\left[\frac{g}{\sqrt{2}}W_{\mu}^{-}\ell_{L,\alpha}^{\dagger}\overline{\sigma}^{\mu}\nu_{L,\alpha}+{\rm h.c.}\right]+\frac{g}{2c_{W}}Z_{\mu}\nu_{L,\alpha}^{\dagger}\overline{\sigma}^{\mu}\nu_{L,\alpha}+{g_{R}Q_{R,j}}Z'_{\mu}\nu_{R,j}^{\dagger}\overline{\sigma}^{\mu}\nu_{R,j},\label{eq:L_chiral}
\end{equation}
where the first three terms are the SM charged and neutral current
interactions, and $\ell_{L}$ denotes left-handed charged leptons.
Therefore, in the mass basis, after performing the basis transformation,
we obtain: 
\begin{equation}
{\cal L}\supset\left[(G_{W})^{\alpha j}W_{\mu}^{-}\ell_{L,\alpha}^{\dagger}\overline{\sigma}^{\mu}\nu_{j}+{\rm h.c.}\right]+(G_{Z})^{ij}Z_{\mu}\nu_{i}^{\dagger}\overline{\sigma}^{\mu}\nu_{j}+(G_{R})^{ij}Z'_{\mu}\nu_{i}^{\dagger}\overline{\sigma}^{\mu}\nu_{j},\label{eq:m-11}
\end{equation}
where 
\begin{equation}
G_{Z}=\frac{g}{2c_{W}}U^{\dagger}\left(\begin{array}{cc}
I_{3\times3}\\
 & 0_{n\times n}
\end{array}\right)U,\ G_{R}=g_{R}U^{\dagger}\left(\begin{array}{cc}
0_{3\times3}\\
 & Q_{R}
\end{array}\right)U,\ \label{eq:GZGR}
\end{equation}
\begin{equation}
G_{W}=\frac{g}{\sqrt{2}}\left(\begin{array}{cc}
I_{3\times3} & 0_{3\times n}\end{array}\right)U.\label{eq:GW}
\end{equation}
Here $Q_{R}={\rm diag}(Q_{R,1},\ Q_{R,2},\ \cdots)$, 
$I_{3\times 3}$ is an identity matrix,  and $0_{x\times y}$ is a zero matrix.

Notice that some products of the above matrices are zero:
\begin{eqnarray}
G_{Z}G_{R} & = & G_{R}G_{Z}=0,\label{eq:GG0}\\
G_{W}G_{R} & = & G_{R}G_{W}^{\dagger}=0.\label{eq:GG0-1}
\end{eqnarray}
The above results, which will be used in our loop calculations to cancel UV divergences, have been previously derived in Ref.~\cite{Pilaftsis:1991ug}.

\section{Loop-induced couplings  of $Z'$\label{sec:loop_coupling}}

\begin{figure}[t]
\centering

\includegraphics[width=0.7\textwidth]{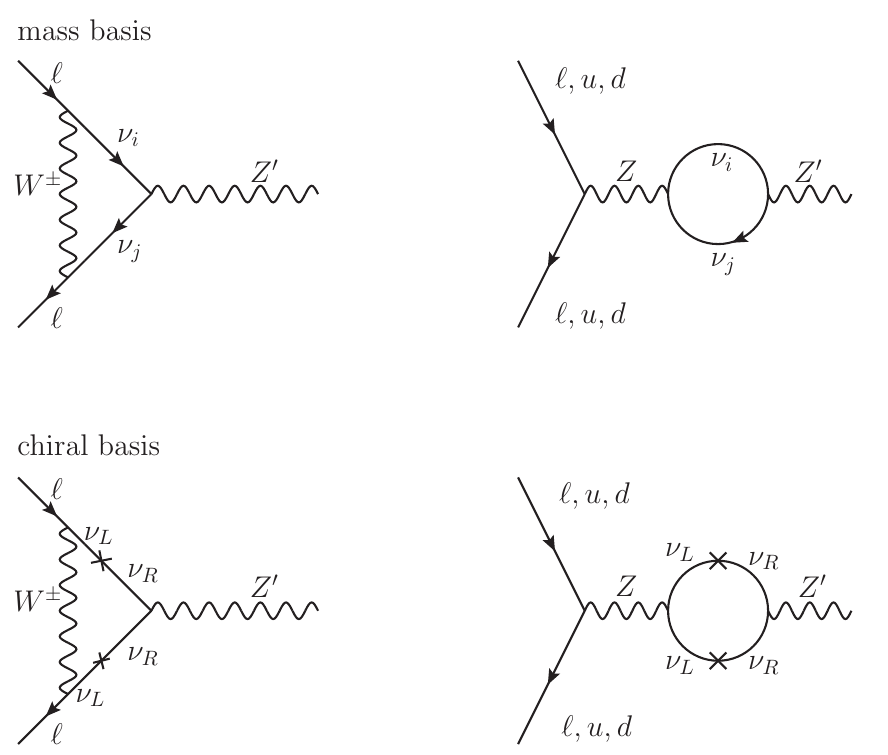}

\caption{ Loop-induced $Z'$ couplings to charged fermions in the mass basis
(upper panels) and in the chiral basis (lower panels). \label{fig:WZ}}
\end{figure}

At tree level, the $\nu_{R}$-philic $Z'$ does not directly couple to charged leptons or quarks.
At the one-loop level, there are loop-induced couplings of $Z'$ generated by the diagrams shown in Fig.~\ref{fig:WZ}.

In the upper and lower panels, we present diagrams in the mass and
chiral bases, respectively. The two descriptions are physically equivalent. The
diagrams in the chiral basis imply that the loop-induced couplings
are proportional to $m_{D}^{2}$, due to the two necessary mass insertions
on the neutrino lines. Although in the mass basis this conclusion
is not evident, technically our calculations are performed using the diagrams
in the upper panel because of properly defined propagators. 
 
Throughout this work, we work in the unitarity gauge so that diagrams
involving Goldstone bosons can be disregarded. The detailed calculations
are presented in Appendix~\ref{appendix:A}.  The result for a single
$W^{\pm}$ diagram with neutrino mass eigenstates $\nu_{i}$ and $\nu_{j}$
running in the loop reads:
\begin{equation}
i\mathcal{M}_{W}^{ij}=i\frac{G_{W}^{\beta j}(G_{R}^{ij}G_{W}^{\alpha i})^{*}}{16\pi^{2}}\mathcal{F}(m_{i},m_{j})\ \overline{u(p_{1})}\gamma^{\mu}P_{L}u(p_{2})\epsilon_{\mu}(q),\label{eq:W}
\end{equation}
where $\overline{u(p_{1})}$ and $u(p_{2})$ denote the two external
fermion states, $\epsilon_{\mu}(q)$ is the polarization vector of
$Z'_{\mu}$, and 
\begin{eqnarray}
\mathcal{F}(m_{i},m_{j})\  & \approx & \ \frac{3}{2}+\frac{m_{j}^{4}\log(m_{j}^{2}/m_{W}^{2})-m_{i}^{4}\log(m_{i}^{2}/m_{W}^{2})}{(m_{i}^{2}-m_{j}^{2})\ m_{W}^{2}}\nonumber \\
 &  & +\ \frac{(m_{i}^{2}+m_{j}^{2})}{m_{W}^{2}}\left[\frac{1}{\epsilon}+1+\log\left(\frac{\mu^{2}}{m_{W}^{2}}\right)\right].\label{eq:FW}
\end{eqnarray}
We have adopted dimensional regularization in the loop calculation
so the loop integral takes the generalized measure $\frac{d^{4}k}{(2\pi)^{4}}\rightarrow\mu^{2\epsilon}\frac{d^{d}k}{(2\pi)^{d}}$
with $d=4-2\epsilon$, which defines $\mu$ and $\epsilon$ in Eq.~\eqref{eq:FW}.

Note that for each single diagram in the mass basis, the result is
UV divergent. However, when we sum over $i$ and $j$, the UV divergence
cancels out. This can be seen as follows:
\begin{equation}
\sum_{ij}\frac{1}{\epsilon}(m_{i}^{2}+m_{j}^{2})G_{W}^{\beta j}(G_{R}^{ij}G_{W}^{\alpha i})^{*}=\frac{1}{\epsilon}G_{W}M_{d}^{2}G_{R}^{\dagger}G_{W}^{\dagger}+\frac{1}{\epsilon}G_{W}G_{R}^{\dagger}M_{d}^{2}G_{W}^{\dagger}=0,\label{eq:x}
\end{equation}
where $M_{d}^{2}\equiv{\rm diag}(m_{1}^{2},\ m_{2}^{2},\ m_{3}^{2},\ \cdots)$
and in the second step we have used Eq.~\eqref{eq:GG0-1}.  Eq.~\eqref{eq:x}
implies that we can safely ignore the second line in Eq.~\eqref{eq:FW},
as long as Eq.~\eqref{eq:GG0-1} holds. For a similar reason ($G_{W}G_{R}^{\dagger}G_{W}^{\dagger}=0$),
the constant term $\frac{3}{2}$ can also be ignored.

For the $Z$ diagram, we have a similar amplitude for each single
diagram. In the soft-scattering limit ($q\rightarrow0$), we find
\begin{equation}
    i\mathcal{M}_{Z}^{ij}=    -\frac{igQ_{Z}^{(f)}G_{Z}^{ij}(G_{R}^{ij})^{*}}{16\pi^{2}c_{W}m_{Z}^{2}}\mathcal{F}_{2}(m_{i},m_{j})\overline{u(p_{1})}\gamma^{\mu}P_{L/R}u(p_{2})\epsilon_{\mu}(q),
\label{eq:Z}
\end{equation}
where $f=\ell_{L/R}$, $u_{L/R}$, or $d_{L/R}$; and $Q_{Z}^{(f)}$ is the $Z$ charge
of $f$, defined in the way that the $Z$-$f$-$f$ coupling can be
written as $gQ_{Z}^{(f)}/c_{W}$. The specific values of $Q_{Z}^{(f)}$
used in this work are listed in Tab.~\ref{table:1}.
\begin{table}
    \centering
    \begin{tabular}{ c c c c c c c c} 
    \hline
    \hline
    $f $ & $\nu_{L}$ & $e_L$ & $u_L$ & $d_L$ & $e_R$ & $u_R$ & $d_R$ \\[1mm] 
    \hline
    $Q_{Z}^{(f)}$ & $\frac{1}{2}$ & $-\frac{1}{2}+s_W^2$ & $\frac{1}{2}-\frac{2}{3}s_W^2$ & $-\frac{1}{2}+\frac{1}{3}s_W^2$ & $s_W^2$ & $-\frac{2}{3}s_W^2$ & $\frac{1}{3}s_W^2$\\[1mm]  
    \hline
    \hline
    \end{tabular}
    \caption{The values of $Q_{Z}^{(f)}$ used in this work.}
    \label{table:1}
\end{table}
The $\mathcal{F}_{2}$ function reads:
\begin{equation}
\mathcal{F}_{2}(m_{i},m_{j})\ \approx\frac{m_{j}^{4}\log(m_{j}^{2})-m_{i}^{4}\log(m_{i}^{2})}{(m_{i}^{2}-m_{j}^{2})\ }+\ (m_{i}^{2}+m_{j}^{2})\left[\frac{1}{\epsilon}+\frac{1}{2}+\log\mu^{2}\right].\label{eq:FZ}
\end{equation}
Once again, we can see that the UV part cancels out during the summation
of $i$ and $j$ because 
\begin{equation}
\sum_{ij}\frac{1}{\epsilon}(m_{i}^{2}+m_{j}^{2})G_{Z}^{ij}(G_{R}^{ij})^{*}=\frac{1}{\epsilon}{\rm Tr}\left[M_{d}^{2}G_{Z}G_{R}^{\dagger}+G_{Z}M_{d}^{2}G_{R}^{\dagger}\right]=0.\label{eq:x-3}
\end{equation}
Hence only the first term in Eq.~\eqref{eq:FZ} needs to be taken
into account.

Summing over $i$ and $j$ in Eq.~\eqref{eq:W}, we obtain the following
effective coupling generated by the loop diagrams:
\begin{equation}
{\cal L}_{{\rm eff}}=\left[g_{{\rm eff},W}\ell_{L,\beta}^{\dagger}\overline{\sigma}^{\mu}\ell_{L,\alpha}+g_{{\rm eff},Z}f^{\dagger}\overline{\sigma}^{\mu}f\right]Z'_{\mu},\label{eq:x-1}
\end{equation}
where
\begin{eqnarray}
g_{{\rm eff},W} & = & \sum_{ij}\frac{G_{W}^{\beta j}(G_{R}^{ij}G_{W}^{\alpha i})^{*}}{16\pi^{2}}\frac{m_{j}^{4}\log(m_{j}^{2}/m_{W}^{2})-m_{i}^{4}\log(m_{i}^{2}/m_{W}^{2})}{(m_{i}^{2}-m_{j}^{2})\ m_{W}^{2}},\label{eq:x-2}\\
g_{{\rm eff},Z} & = & \sum_{ij}\frac{ gQ_{Z}^{(f)} \ G_{Z}^{ij}(G_{R}^{ij})^{*}}{16\pi^{2}c_W}\frac{m_{i}^{4}\log(m_{i}^{2})-m_{j}^{4}\log(m_{j}^{2})}{(m_{i}^{2}-m_{j}^{2})m_{Z}^{2}\ }.\label{eq:x-5}
\end{eqnarray}

\subsection{Example A: $1\ \nu_{L}+1\ \nu_{R}$}
First, let us consider the simplest case that there are only one $\nu_{L}$ and one $\nu_{R}$. The neutrino mass matrix $M_\nu$ for the case can be diagonalized by a $2\times2$ unitary matrix 
\begin{equation}
U^{T}\left(\begin{array}{cc}
0 & m_{D}\\
m_{D} & M_{R}
\end{array}\right)U=\left(\begin{array}{cc}
m_{1}& 0\\
0 & m_{4}
\end{array}\right).\label{eq:U1}
\end{equation}
This unitary matrix can be parametrized as follows 
\begin{equation}
U=\left(\begin{array}{cc}
-i c_\theta & s_\theta\\
i s_\theta & c_\theta
\end{array}\right),\ \ \theta=\arctan \left(\sqrt{\frac{m_1}{m_4}}\right), \label{eq:u1}
\end{equation}
where $c_\theta=\cos\theta$ and $s_\theta=\sin\theta$. Substituting  the explicit form of $U$ in Eqs.~\eqref{eq:GZGR} and \eqref{eq:GW}, we obtain
\begin{equation}
G_{Z}=\frac{g}{2c_{W}}\left(\begin{array}{cc}
c_\theta^2 & i c_\theta s_\theta\\
-i c_\theta s_\theta & s_\theta^2
\end{array}\right),\ G_{R}=g_{R}\left(\begin{array}{cc}
s_\theta^2 & -i c_\theta s_\theta\\
i c_\theta s_\theta & c_\theta^2
\end{array}\right),\ \label{eq:GZGR1}
\end{equation}
\begin{equation}
G_{W}=\frac{g}{\sqrt{2}}\left(\begin{array}{cc}
-i c_\theta & s_\theta\end{array}\right).\label{eq:GW1}
\end{equation}
We can now perform the summation in Eqs.~\eqref{eq:x-2}-\eqref{eq:x-5}. Expanding the result as a Taylor series in $s_\theta$ (assuming $s_\theta \ll 1$) and only retaining the dominant contribution, we obtain 
\begin{eqnarray}
g_{{\rm eff},W} & = & -\frac{g^2 m^2s_\theta^2}{32 \pi^2 m_W^2}g_R,\label{eq:geffw1}\\
g_{{\rm eff},Z} & = & Q_{Z}^{(f)}\frac{\ g^2 m^2s_\theta^2}{32 \pi^2 m_Z^2 c_W^2}g_R\,,\label{eq:geffz1}
\end{eqnarray}
where
\begin{equation}
    m_1=ms_{\theta}^2, \ m_4=mc_{\theta}^2.
\end{equation}
Note that for $m_1 \ll m_4$,
\begin{equation}
m^2 s^2_\theta \simeq m_1 \,m_4 = m_D^2.
\end{equation}
Using $G_F=\frac{\sqrt{2}g^2}{8m_W^2}$, we can rewrite Eqs.~\eqref{eq:geffw1}-\eqref{eq:geffz1} as
\begin{eqnarray}
g_{{\rm eff},W} & = & -\frac{\sqrt{2}\, G_F\, m_D^2}{8 \pi^2 }g_R,\label{eq:geffw1f}\\
g_{{\rm eff},Z} & = &Q_{Z}^{(f)}\frac{\sqrt{2}\, G_F\, m_D^2}{8 \pi^2 }g_R .\label{eq:geffz1f}
\end{eqnarray}

\subsection{Example B: $1\ \nu_{L}+2\ \nu_{R}$ with opposite charges \label{sub:modelB}}
In this example, we construct a UV-complete model with one $\nu_{L}$ and two $\nu_{R}$ which have opposite $U(1)_R$ 
charges so that the model is anomaly free. The off-diagonal Majorana mass term does not violate the $U(1)_R$  symmetry and 
the Dirac mass term is generated by a new Higgs doublet $H'$ that is charged under $U(1)_R$. 
The $U(1)_R$ 
charges are assigned as follows: 
\begin{equation}
Q_{R}(\nu_{R,1})=+1,\
Q_{R}(\nu_{R,2})=-1,\ 
Q_{R}(H')= -1,
\end{equation} 
which leads to the following terms that fully respect the $U(1)_R$ symmetry:
\begin{equation}
   {\cal L}\supset y_\nu\widetilde{H'}^\dagger L\nu_{R_1} + \frac{M_R}{2}\nu_{R_1}\nu_{R_2}+{\rm h.c.},
\end{equation}
where $\widetilde{H'}\equiv i\sigma_2 (H')^*$.
After spontaneous symmetry breaking, $H'$ acquires a vacuum expectation value : $\langle H'\rangle=(0,\ v')^T/\sqrt{2}$, leading to
\begin{equation}
   {\cal L} \supset m_D \nu_L\nu_{R_1} + \frac{M_R}{2}\nu_{R_1}\nu_{R_2}+{\rm h.c.}
\end{equation}
Here $m_D=y_\nu {v'}/{\sqrt{2}}$. 
The neutrino mass matrix for this case can be diagonalized by a $3\times3$ unitary matrix: 
\begin{equation}
U^{T}\left(\begin{array}{ccc}
0 & m_{D} & 0\\
m_{D} & 0 & M_{R} \\
0 & M_R & 0 
\end{array}\right)U=\left(\begin{array}{ccc}
m_{1}& 0 & 0\\
0 & m_{4} & 0 \\
0 & 0 & m_{5} 
\end{array}\right).\label{eq:UMU}
\end{equation}
The texture of the mass matrix 
on the left-hand side of Eq.~\eqref{eq:UMU} 
leads to $m_1=0$ and $m_4=m_5$, which is evident from its vanishing trace and determinant. 
This feature has been often considered in the literature on $\nu_R$ signals at the LHC---see e.g.~\cite{Kersten:2007vk,Deppisch:2015qwa} and references therein.
The $3\times3$ unitary matrix can be parametrized as follows 
\begin{equation}
U=\left(\begin{array}{ccc}
-c_\theta & \frac{is_\theta}{\sqrt{2}} & \frac{s_\theta}{\sqrt{2}}\\
0 & \frac{-i}{\sqrt{2}} & \frac{1}{\sqrt{2}}\\
s_\theta & \frac{ic_\theta}{\sqrt{2}} & \frac{c_\theta}{\sqrt{2}}
\end{array}\right),\ \ \theta=\arctan \left({\frac{m_D}{M_R}}\right). \label{eq:u2}
\end{equation}
Using this form of $U$ in Eqs.~\eqref{eq:GZGR} and \eqref{eq:GW}, we obtain
\begin{equation}
G_{Z}=\frac{g}{2c_{W}}\left(\begin{array}{ccc}
c_\theta^2 & \frac{-i c_\theta s_\theta}{\sqrt{2}} & \frac{- c_\theta s_\theta}{\sqrt{2}}\\
\frac{i c_\theta s_\theta}{\sqrt{2}} & \frac{s_\theta^2}{2} & \frac{-i s_\theta^2}{{2}}\\
\frac{- c_\theta s_\theta}{\sqrt{2}} & \frac{i s_\theta^2}{{2}} & \frac{s_\theta^2}{{2}}
\end{array}\right),\ G_{R}=g_{R}\left(\begin{array}{ccc}
-s_\theta^2 & \frac{-i c_\theta s_\theta}{\sqrt{2}} & \frac{- c_\theta s_\theta}{\sqrt{2}}\\
\frac{i c_\theta s_\theta}{\sqrt{2}} & \frac{s_\theta^2}{2} & \frac{i(1+ c_\theta^2)}{{2}}\\
\frac{- c_\theta s_\theta}{\sqrt{2}} & \frac{-i(1+ c_\theta^2)}{{2}} & \frac{s_\theta^2}{{2}}
\end{array}\right),\ \label{eq:GZGR2}
\end{equation}
\begin{equation}
G_{W}=\frac{g}{\sqrt{2}}\left(\begin{array}{ccc}
-c_\theta & \frac{is_\theta}{\sqrt{2}} & \frac{s_\theta}{\sqrt{2}}\end{array}\right).\label{eq:GW2}
\end{equation}
We can now perform the summation in Eqs.~\eqref{eq:x-2}-\eqref{eq:x-5}. Expanding the result as a Taylor series in $s_\theta$ (assuming $s_\theta \ll 1$) and only retaining the dominant contribution, we obtain
\begin{eqnarray}
g_{{\rm eff},W} & = & \frac{g^2c_\theta^2m^2s_\theta^2}{32 \pi^2 m_W^2}g_R,\label{eq:geffw2}\\
g_{{\rm eff},Z} & = & -Q_{Z}^{(f)}\frac{\ g^2c_\theta^2m^2s_\theta^2}{32 \pi^2 m_Z^2 c_W^2}g_R\,,\label{eq:geffz2}
\end{eqnarray}
where $m\equiv \sqrt{m_D^2+M_R^2}$ and
\begin{equation}
    m_D=ms_{\theta}, \ M_R=mc_{\theta}.
\end{equation}
Expressing the results in terms of $G_F$ and assuming $s_{\theta} \ll 1$, we obtain
\begin{eqnarray}
g_{{\rm eff},W} & = & \frac{\sqrt{2}\ G_F\ m_D^2}{8 \pi^2 }g_R,\label{eq:geffw2f}\\
g_{{\rm eff},Z} & = & -Q_{Z}^{(f)}\frac{\sqrt{2}\ G_F\ m_D^2}{8 \pi^2 }g_R .\label{eq:geffz2f}
\end{eqnarray}
We comment here that the above UV-complete and anomaly-free model built on $1\,\nu_{L}+2\,\nu_{R}$ can be straightforwardly generalized to $3\,\nu_{L}+2n\,\nu_{R}$ where half of the right-handed neutrinos have opposite $U(1)_R$ charges to the other half. Such a generalization can accommodate the realistic three-neutrino mixing measured in neutrino oscillation experiments.

\subsection{Example C: $3\ \nu_{L}+3\ \nu_{R}$ with diagonal $M_{R}$}

The most general case with three $\nu_{L}$ and an arbitrary number of $\nu_R$ is complicated and often impossible to be computed analytically. Here we consider an analytically calculable example   with $3\,\nu_{L}+3\,\nu_{R}$ and the following form of the neutrino mass matrix:
\begin{equation}
\left(\begin{array}{cc}
0_{3\times3} & m_{D}\\
m_{D}^{T} & M_{R}
\end{array}\right)=\left(\begin{array}{cc}
U_{L}^{*} & 0\\
 & I_{3\times3}
\end{array}\right)\left(\begin{array}{cc}
0_{3\times3} & m_{D}^{(d)}\\
m_{D}^{(d)} & M_{R}^{(d)}
\end{array}\right)\left(\begin{array}{cc}
U_{L}^{\dagger} & 0\\
 & I_{3\times3}
\end{array}\right),\label{eq:x-6}
\end{equation}
\[
m_{D}^{(d)}={\rm diag}(m_{D1},\ m_{D2},\ m_{D3}),\ M_{R}^{(d)}={\rm diag}(M_{R1},\ M_{R2},\ M_{R3}),
\]
where $U_L$ is a $3\times 3$ unitary matrix.  Eq.~\eqref{eq:x-6} is not the most general form, but at least it
can accommodate the realistic low-energy neutrino mixing responsible
for neutrino oscillation.
\par The mass matrix in this case can be diagonalized by a $6\times6$ unitary matrix:
\begin{equation}
U'^{T}\left(\begin{array}{cc}
0_{3\times3} & m_{D}^{(d)}\\
m_{D}^{(d)} & M_{R}^{(d)}
\end{array}\right)U'={\rm diag}(m_{1}s_{\theta 1}^2,\, m_{2}s_{\theta 2}^2,\, m_{3}s_{\theta 3}^2,\, m_{1}c_{\theta 1}^2,\, m_{2}c_{\theta 2}^2,\, m_{3}c_{\theta 3}^2),
\label{eq:U3}
\end{equation}
where $(s_{\theta i},\, c_{\theta i})\equiv (\sin {\theta_i},\, \cos {\theta_i})$ and 
\begin{equation}
m_i = \sqrt{4m_{Di}^2+M_{Ri}^2},\ \theta_i=\frac{1}{2}\arctan \left({\frac{2\,m_{Di}}{M_{Ri}}}\right).
\end{equation}
The unitary matrix $U'$ can be parametrized as follows 
\begin{equation}
U'=\left(\begin{array}{cc}
-iC_{\theta} & S_{\theta}\\
iS_{\theta} & C_{\theta}
\end{array}\right), \label{eq:u3}
\end{equation}
where
\begin{equation}
    C_{\theta}={\rm diag}(c_{\theta 1},c_{\theta 2},c_{\theta 3}),\ S_{\theta}={\rm diag}(s_{\theta 1},s_{\theta 2},s_{\theta 3}).
\end{equation}
Thus, the final unitary matrix $U$ that diagonalizes the original mass matrix is given by
\begin{equation}
U=\left(\begin{array}{cc}
U_{L} & 0\\
 & I_{3\times3}
\end{array}\right)\left(\begin{array}{cc}
-iC_{\theta} & S_{\theta}\\
iS_{\theta} & C_{\theta}
\end{array}\right)=\left(\begin{array}{cc}
-iU_{L}C_{\theta} & U_{L}S_{\theta}\\
iS_{\theta} & C_{\theta}
\end{array}\right).
\end{equation}
Substituting it in Eqs.~\eqref{eq:GZGR} and \eqref{eq:GW}, we obtain
\begin{equation}
G_{Z}=\frac{g}{2c_{W}}\left(\begin{array}{cc}
C_\theta^2 & i C_\theta S_\theta\\
-i C_\theta S_\theta & S_\theta^2
\end{array}\right),
\ G_{R}=g_{R}Q_R\left(\begin{array}{cc}
S_\theta^2 & -i C_\theta s_\theta\\
i C_\theta S_\theta & C_\theta^2
\end{array}\right),\ \label{eq:GZGR3}
\end{equation}
\begin{equation}
G_{W}=\frac{g}{\sqrt{2}}U_{L}\left(\begin{array}{cc}
-i C_\theta & S_\theta\end{array}\right),\ Q_R={\rm diag}(Q_{R1},Q_{R2},Q_{R3}).\label{eq:GW3}
\end{equation}
Next, we perform the summation in Eqs.~\eqref{eq:x-2}-\eqref{eq:x-5}, expand the result in $s_{\theta i}$,  and retain the dominant contribution.
The final result reads
\begin{eqnarray}
g_{{\rm eff},W}^{\alpha\beta} & = & \sum_i -U_L^{\beta i}(U_L^{\alpha i})^*Q_{Ri}\frac{\sqrt{2}\ G_F\ m_{Di}^2}{8 \pi^2 }g_R,\label{eq:geffw3f}\\
g_{{\rm eff},Z} & = & \sum_i Q_{Z}^{(f)}\ Q_{Ri}\frac{\sqrt{2}\ G_F\ m_{Di}^2}{8 \pi^2 }g_R .\label{eq:geffz3f}
\end{eqnarray}
In the approximation that the $\nu_L$-$\nu_R$ mixing is small, the $3\times3$ unitary matrix $U_L$ is almost identical to the PMNS matrix.
Due to the presence of off-diagonal entries in $U_L$,  $g_{{\rm eff},W}^{\alpha\beta}$ is generally not flavor diagonal and might lead to observable lepton flavor violation, which will be discussed in Sec.~\ref{sec:pheno}.

\section{Dark photon masses and technical naturalness \label{sec:mass}}

In this section, we argue that despite being a free parameter, the
mass of the $\nu_{R}$-philic dark photon $m_{Z'}$ is potentially
related to the gauge coupling according to 't Hooft's technical naturalness \cite{tHooft:1979rat}. Generally speaking, from the consideration of model building and
the stability of $m_{Z'}$ under loop corrections, we expect that
$m_{Z'}$ is related to $g_{R}$ by
\begin{equation}
m_{Z'}\gtrsim g_{R}\Lambda_{{\rm breaking}},\label{eq:x-7}
\end{equation}
where $\Lambda_{{\rm breaking}}$ stands for the symmetry breaking
scale of $U(1)_{R}$. Although without UV completeness we cannot have
a more specific interpretation of Eq.~\eqref{eq:x-7}, we would like
to discuss a few examples to show how $m_{Z'}$ is related to $g_{R}$.

First, let us consider that both $m_{Z'}$ and $M_{R}$ arise from
a scalar singlet $\phi$ charged under $U(1)_{R}$ with $\langle\phi\rangle=v_{R}\neq0$.
This leads to $m_{Z'}\sim g_{R}v_{R}$ and $M_{R}\sim y_{R}v_{R}$
where $y_{R}$ is the Yukawa coupling of $\phi$ to $\nu_{R}$. In
this case, we consider $v_{R}$ as the symmetry breaking scale $\Lambda_{{\rm breaking}}$
so the tree-level relation $m_{Z'}\sim g_{R}\,v_{R}$ is compatible
with Eq.~\eqref{eq:x-7}. The Yukawa coupling has an upper bound from
perturbativity, $y_{R}\lesssim4\pi$, which implies that $m_{Z'}/M_{R}\sim g_{R}/y_{R}\gtrsim4\pi g_{R}$,
or
\begin{equation}
m_{Z'}^{2}\gtrsim\frac{g_{R}^{2}}{16\pi^{2}}M_{R}^{2}.\label{eq:x-8}
\end{equation}
In the absence of a specific symmetry breaking mechanism, we can also
obtain Eq.~\eqref{eq:x-8} purely from loop corrections to $m_{Z'}$.
If $M_{R}$ breaks the $U(1)_{R}$ symmetry, the $Z'$-$Z'$ vacuum
polarization amplitude generated by a $\nu_{R}$ loop is $\Pi^{\mu\nu}(q^{2})\sim\frac{g_{R}^{2}}{16\pi^{2}}\left[{\cal O}(M_{R}^{2})g^{\mu\nu}+{\cal O}(1)q^{\mu}q^{\nu}\right]$,
which implies that the loop correction to $m_{Z'}^2$ is of the order
of $\frac{g_{R}^{2}}{16\pi^{2}}M_{R}^{2}$. Therefore, to make the
theory technically natural, the physical mass should not be lower
than the loop correction.

Note, however, that Eq.~\eqref{eq:x-8} is based on the assumption
that $M_{R}$ breaks the $U(1)_{R}$ symmetry. If all the Majorana
mass terms fully respect  $U(1)_{R}$, such as Example B in Sec.~\ref{sec:loop_coupling},
then the symmetry breaking scale can be lower, e.g., determined by
$m_{D}$. Indeed, for the UV complete model in Example B, the symmetry
breaking scale is determined by the VEV of the new Higgs doublet $H'$
so at tree level we have $m_{Z'}\sim g_{R}\langle H'\rangle$ and
$m_{D}\sim y_{D}\langle H'\rangle$. Then using the perturbativity
bound on $y_{D}$, we obtain
\begin{equation}
m_{Z'}^{2}\gtrsim\frac{g_{R}^{2}}{16\pi^{2}}m_{D}^{2}.\label{eq:x-9}
\end{equation}
Finally, we comment on the possible mass correction from $Z$-$Z'$
mixing. According to the calculation in Appendix~\ref{appendix:A}, the
vacuum polarization diagram leads to mass mixing between $Z$ and $Z'$:
\begin{equation}
{\cal L}_{ZZ'\thinspace{\rm mass}}=\frac{1}{2}(Z,\ Z')^{\mu}\left(\begin{array}{cc}
m_{Z_{0}}^{2} & m_{X}^{2}\\
m_{X}^{2} & m_{Z'_{0}}^{2}
\end{array}\right)\left(\begin{array}{c}
Z\\
Z'
\end{array}\right)_{\mu},
\end{equation}
where $m_{Z_{0}}$ and $m_{Z'_{0}}$ denote tree-level masses and
\begin{equation}
m_{X}^{2}=\frac{g_{R}Q_{R}\ g}{64\pi^{2}\cos{\theta_{W}}}m_{D}^{2}.
\end{equation}
Here $m_{X}^{2}$ causes $Z-Z'$ mixing and the mixing angle is roughly
$\frac{m_{X}^{2}}{|m_{Z_{0}}^{2}-m_{Z'_{0}}^{2}|}$, which must be
small. Otherwise, the SM neutral current would be significantly modified
and become inconsistent with electroweak precision data. Taking the
approximation $m_{X}^{2}\ll|m_{Z_{0}}^{2}-m_{Z'_{0}}^{2}|$, we 
obtain
\begin{equation}
m_{Z}^{2}\simeq m_{Z_{0}}^{2}+\frac{m_{X}^{4}}{(m_{Z_{0}}^{2}-m_{Z'_{0}}^{2})},\ \ m_{Z'}^{2}\simeq m_{Z'_{0}}^{2}-\frac{m_{X}^{4}}{(m_{Z_{0}}^{2}-m_{Z'_{0}}^{2})}.\label{eq:x-10}
\end{equation}
Hence we conclude that the mass correction from $Z$-$Z'$ mixing
is 
\begin{equation}
\delta m_{Z'}^{2}\sim\frac{g_{R}^{2}}{(64\pi^{2})^{2}}\frac{m_{D}^{4}}{|m_{Z_{0}}^{2}-m_{Z'_{0}}^{2}|},\label{eq:x-11}
\end{equation}
where we have neglected some ${\cal O}(1)$ quantities. This mass
correction is generally smaller than the right-hand side of Eq.~\eqref{eq:x-9}
because $m_{D}$ cannot be much above the electroweak scale.

To summarize, here we draw 
a less model-dependent
conclusion that without
fine-tuning, the $\nu_{R}$-philic dark photon mass is expected to
be above the lower bound in Eq.~\eqref{eq:x-8} or Eq.~\eqref{eq:x-9},
depending on whether $M_{R}$ breaks the $U(1)_{R}$ symmetry or not,
respectively.

\section{Phenomenology}\label{sec:pheno}

\begin{figure}[t]
\centering

\includegraphics[width=0.85\textwidth]{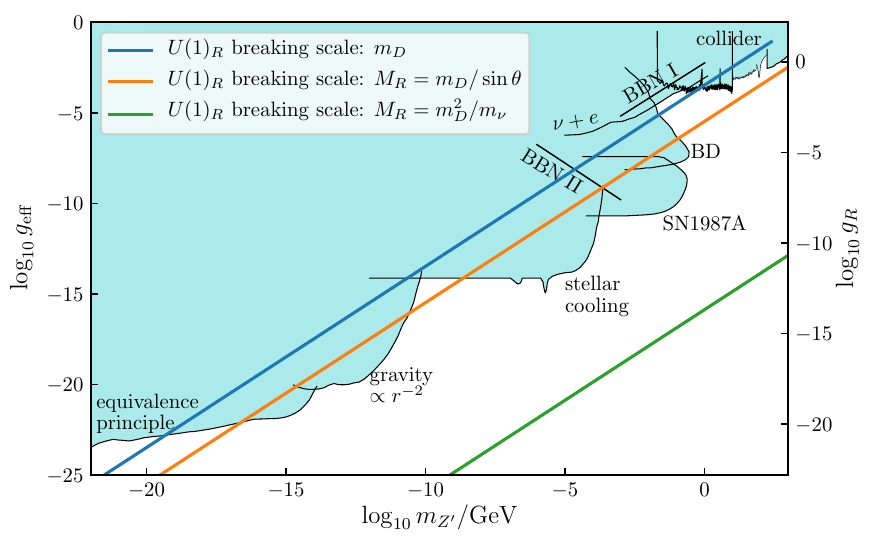}

\caption{The $\nu_{R}$-philic dark photon confronted with known experimental constraints. 
Here $g_{\rm eff}$ is the loop-induced coupling of $Z'$ to electrons.
The quark couplings are of the same order of magnitude as $g_{\rm eff}$ and we have ignored the difference between them when recasting constraints on quark couplings.
  The theoretically favored values of $g_{\rm eff}$ are below the solid blue, orange, or green lines, assuming $U(1)_R$ breaks at the scale of $m_D=246$ GeV, $M_R=24.6$ TeV, or $M_R\sim 10^{14}$ GeV (Type I seesaw), respectively. 
The collider bound consists of  BaBar, LHCb, LEP, and
LHC 8 TeV limits---see the text or Fig.~\ref{fig:bound-1} for more
details. 
\label{fig:bound}}
\end{figure}

\begin{figure}[t]
\centering

\includegraphics[width=0.85\textwidth]{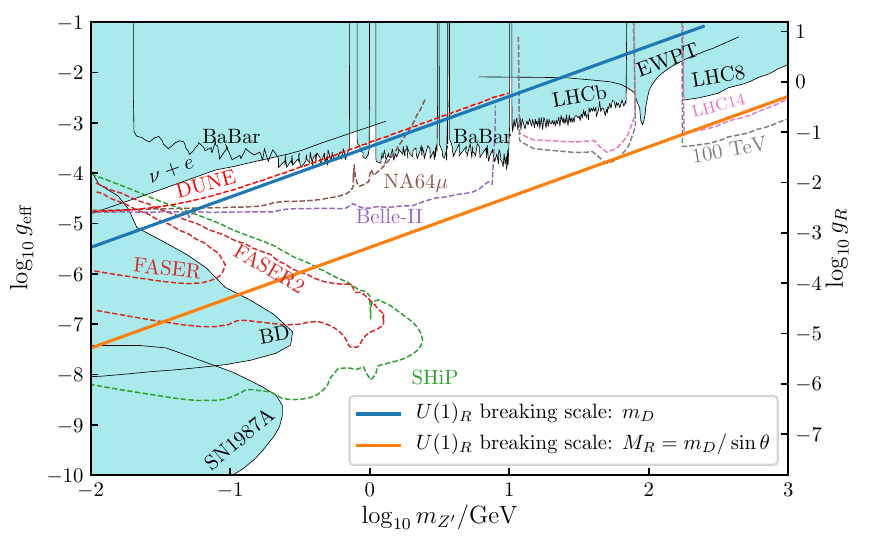}

\caption{Sensitivity of future experiments (SHiP, FASER, Belle-II) on the $\nu_{R}$-philic dark photon. 
Here $g_{\rm eff}$ is the loop-induced coupling of $Z'$ to electrons.
The quark couplings are of the same order of magnitude as $g_{\rm eff}$ and we have ignored the difference between them when recasting constraints on quark couplings.
The theoretically favored values of $g_{\rm eff}$ is below the solid blue or orange lines, assuming $U(1)_R$ breaks at the scale of $m_D=246$ GeV or $M_R=24.6$ TeV, respectively.
\label{fig:bound-1}}
\end{figure}

\noindent In the previous two sections, we have derived the loop-induced
couplings and also  argued that from technical naturalness there is
a lower bound on the dark photon mass.  The results indicate the
theoretically favored regime of the mass and the couplings. Therefore,
to address the question of how dark the $\nu_{R}$-philic dark photon
would be, we shall inspect whether and to what extent the theoretically
favored regime could be probed by current and future experiments.

In our model, there are effective couplings to both leptons and quarks
with comparable strengths. So the experimental constraints on this
model are very similar to those on the $B-L$ model\footnote{See e.g.~Fig.~8 in \cite{Harnik:2012ni}, Fig.~3 in \cite{Heeck:2014zfa}, and Fig.~13
in \cite{Bauer:2018onh}}. Below we discuss a variety of known bounds that could be important
for the $\nu_{R}$-philic dark photon. An overview of existing bounds
is presented  in Fig.~\ref{fig:bound}, and the prospect of upcoming
experiments in Fig.~\ref{fig:bound-1}.

\subsection{Experimental limits}

\subsubsection{Collider searches}

With effective couplings to electrons and quarks, dark photons could
be produced directly in $e^{+}e^{-}$ (BaBar, LEP) and  hadron colliders
(LHC), typically manifesting themselves as resonances in collider
signals. For $m_{Z'}\gtrsim175$ GeV ($t$ quark resonance), LHC data
put the strongest bound via Drell-Yan production and detection of
leptonic final states ($pp\rightarrow Z'\rightarrow\ell^{+}\ell^{-}$).
At lower masses when $m_{Z'}$ is close to the $Z$ pole, electroweak
precision tests (EWPT, including LEP measurement and other electroweak
precision observables) become more important. A dedicated analysis
on LHC and EWPT bounds and future prospects can be found in Ref.~\cite{Curtin:2014cca}.
For $m_{Z'}$ below the $Z$ pole but above 10 GeV, according to the
analyses in \cite{Bauer:2018onh}, the most stringent constraint comes from LHCb
di-muon ($Z'\rightarrow\mu^{+}\mu^{-}$) measurements \cite{Aaij:2017rft}. Below 10 GeV, the BaBar experiment \cite{Lees:2014xha} provides more stringent
constraints via $e^{+}e^{-}\rightarrow\gamma Z'$ where $Z'$ may
or may not decay to visible final states. In Figs.~\ref{fig:bound}
and \ref{fig:bound-1}, we present all aforementioned constraints (for compactness in Fig.~\ref{fig:bound} they are labeled
together as the collider bound). Besides, there is also an indirect LEP bound on four-fermion effective interactions---see Sec.~3.5.2 in Ref.~\cite{Schael:2013ita}.  We find that this bound approximately corresponds to $g_{{\rm eff}}/m_{Z'}\lesssim (4.4\ {\rm TeV})^{-1}$, which is weaker than the aforementioned collider bounds and hence not shown in  Figs.~\ref{fig:bound} and \ref{fig:bound-1}.

\subsubsection{Beam dump and neutrino scattering bounds}

For $1\ {\rm MeV}\lesssim m_{Z'}\lesssim100$ MeV, beam dump (BD)
and neutrino scattering experiments become important. BD experiments
search for dark photons by scattering an electron/proton beam on fixed
targets and looking for dark particles that might be produced and
subsequently decay after the shield to visible particles such as electrons.
A compilation of existing BD bounds from SLAC E141, SLAC E137, Fermilab
E774, Orsay, and KEK experiments can be found in \cite{Andreas:2012mt}.
Note that these BD bounds relies on $Z'\rightarrow e^{+}e^{-}$ decay,
which implies that such bounds do not apply  for $m_{Z'}\lesssim2m_{e}$.
Nonetheless, below 1 MeV there are much stronger bounds from cosmological
and astrophysical observations hence for simplicity we do not show
the invalidity of BD bounds below 1 MeV. The combined BD bound adopted
in this work is  taken from \cite{Heeck:2014zfa}.

The dark photon in our model could contribute to elastic neutrino
scattering by a new neutral-current-like process. Current data from
elastic neutrino-electron (CHARM-II~\cite{Vilain:1993kd,Vilain:1994qy},
TEXONO~\cite{Deniz:2009mu}, GEMMA~\cite{Beda:2010hk}, Borexino~\cite{Bellini:2011rx},
etc.) and neutrino-nucleus (COHERENT \cite{Akimov:2017ade}) scattering
are all well consistent with the SM predictions. By comparing the
results in Refs. ~\cite{Bilmis:2015lja,Farzan:2018gtr,Lindner:2018kjo},
we find that the COHERENT bound is weaker than $\nu+e$ scattering
bounds, among which the most stringent ones come from CHARM-II, TEXONO,
and GEMMA. So the combined result from these experiments is taken
from Ref.~\cite{Lindner:2018kjo} and presented in Figs.~\ref{fig:bound}
and \ref{fig:bound-1}. The future DUNE experiment will be able to further improve the measurement of elastic neutrino
scattering~\cite{Abi:2020kei}. We adopt the DUNE sensitivity from Ref.~\cite{Ballett:2019xoj} and present it in Fig.~\ref{fig:bound-1}.

\subsubsection{Astrophysical and cosmological bounds}

Astrophysical bounds on dark photons are usually derived from energy
loss in celestial bodies such as the sun, red giants, horizontal branch
stars, and supernovae. Dark photons may contribute to stellar energy
loss directly via dark photon free streaming or indirectly via neutrino
production. The enhanced energy loss rate could alter stellar evolution
on the horizontal branch in the Hertzsprung-Russell diagram. This
sets the strongest limit for sub-MeV dark photons \cite{Redondo:2013lna}.
For smaller $m_{Z'}$, there are also similar bounds from the sun
and red giants \cite{Redondo:2013lna}. We adopt a combined bound from Ref.~\cite{Harnik:2012ni}
with energy loss via neutrinos taken into account, and refer to it
as the stellar cooling bound in Fig.~\ref{fig:bound}. 

The observation of SN1987A can be used to set strong limits on the
effective coupling when $m_{Z'}\lesssim{\cal O}(100)$ MeV \cite{Dent:2012mx}.
The resulting bound further excludes the space below BD constraints
by about three orders of magnitude.

In Fig.~\ref{fig:bound}  we also show two bounds derived from the effect of $Z'$
on big bang nucleosynthesis (BBN). The effect of $Z'$ on BBN is two-fold:
if $Z'$ is light and dominantly decays to invisible states, it would
increase the effective number of relativistic dark species $N_{{\rm eff}}$.
We refer to the bound derived from this effect as the BBN II bound.
If $Z'$ is heavy, it decays before neutrino decoupling and does not
contribute to $N_{{\rm eff}}$ directly but the neutrino decoupling
temperature could be modified if $g_{{\rm eff}}^{2}/m_{Z'}^{2}$ is
comparable to $G_{F}$ (referred to as BBN I). Among various studies
on this subject (see e.g.~\cite{Knapen:2017xzo,Dutta:2020jsy,Luo:2020sho,Dutta:2020enk,Luo:2020fdt}),
we adopt the bounds from \cite{Knapen:2017xzo} for the $B-L$ model and
label them as BBN I and BBN II in Fig.~\ref{fig:bound}.

\subsubsection{Charged lepton flavor violation}
The loop-induced couplings do not necessarily conserve lepton flavors, as indicated by Eq.~\eqref{eq:geffw3f}. Note, however, that neither the $W$-diagram nor the $Z$-diagram causes flavor violation in the quark sector. In the presence of flavor-changing couplings of $Z'$ to charged leptons, there are strong constraints from charged lepton flavor violating (CLFV) decay such as $\ell_\alpha\rightarrow\ell_\beta\nu\bar{\nu}$, $\mu\rightarrow 3 e$~\cite{Langacker:2000ju}, $\pi^0\rightarrow e \mu$; from  $\mu\rightarrow e$ conversion in muonic atoms~\cite{Kaulard:1998rb}, and from the non-observation of muonium-antimuonium transitions \cite{Willmann:1998gd}. Constraints from $\ell_{\alpha}\rightarrow\ell_{\beta}\gamma$ are weaker since they arise only from two-loop contributions. We do not include CLFV bounds in Figs.~\ref{fig:bound}  and \ref{fig:bound-1} because such bounds depend on the flavor structure of $m_D$ which in the Casas-Ibarra parametrization~\cite{Casas:2001sr}: $m_{D}=iU_{L}^{*}\sqrt{m_{\nu} }R^{T}\sqrt{M_{R}}$ where $R$ is a complex orthogonal matrix, depends not only on the PMNS matrix $U_{L}$ but also on the $R$ matrix.
The effective flavor-changing  couplings in the presence of non-trivial $R$ are more complicated and we leave them for future work.

\subsubsection{Long-range force searches }

Below 0.1 eV, laboratory tests of gravity and gravity-like forces
provide highly restrictive constraints, including high precision tests
of the inverse-square law (gravity $\propto r^{-2}$)~\cite{Adelberger:2006dh,Adelberger:2009zz}
and of the equivalence principle via torsion-balance experiments~\cite{Wagner:2012ui}
and lunar laser-ranging (LLR) measurements~\cite{Wagner:2012ui,Turyshev:2006gm}.
Besides, measurements of the Casimir effect~\cite{Bordag:2001qi}
could set a limit that is slightly stronger than that from the inverse-square
law when $0.05\lesssim m_{Z'}/{\rm eV}\lesssim0.1$, which is not
presented in Fig.~\ref{fig:bound}. Also not presented here is the
bound from black hole superradiance~\cite{Baryakhtar:2017ngi}, which
would only enter the lower left corner in Fig.~\ref{fig:bound}.
We refer to our previous work~\cite{Xu:2020qek} for more detailed
discussions on the long-range force searches and present only the
dominant constraints from torsion-balance tests of the inverse-square
law and the equivalence principle.
We comment here that neutrino oscillation could also be used to probe long-range forces~\cite{Wise:2018rnb,Bustamante:2018mzu,Smirnov:2019cae,Babu:2019iml} but similar to the aforementioned CLFV bounds, the flavor structure cannot be simply taken into account by the PMNS matrix.
Hence we leave this possibility to future studies.

\subsubsection{Prospect of upcoming experiments}

Future hadron collider searches could significantly improve the experimental
limits  on heavy dark photons by almost one order of magnitude, as
illustrated in Fig.~\ref{fig:bound-1} by the LHC 14 TeV and future
100 TeV collider sensitivity \cite{Curtin:2014cca}. Moreover, several LHC-based
experiments searching for displaced dark photon decays such as FASER \cite{Feng:2017uoz}, MATHUSLA \cite{Chou:2016lxi,Evans:2017lvd}, and CodexB \cite{Gligorov:2017nwh} will improve the BD bound in the low-mass regime.
And the future SHiP experiment \cite{Anelli:2015pba,Alekhin:2015byh} will substantially
broaden the BD bound regarding both the dark photon mass and coupling.
The current BaBar bound may be superseded by future bounds from Belle-II \cite{Abe:2010gxa} and a muon run of NA64 \cite{Banerjee:2016tad,Gninenko:2018tlp}.
Hence a large part of the space that is often considered for dark
photons ($20{\rm MeV}\lesssim m_{Z'}\lesssim10$ GeV and $10^{-8}\lesssim g_{{\rm eff}}\lesssim10^{-3}$)
will be probed by future experiments. Here we selectively present
the sensitivity curves of SHiP, FASER, NA64$\mu$, and Belle-II. Most of them are taken from Ref.~\cite{Bauer:2018onh}, except for the FASER/FASER2 sensitivity which is taken from Ref.~\cite{Ariga:2018uku}.

\subsection{How dark is the $\nu_{R}$-philic dark photon?}

Since the effective coupling $g_{{\rm eff}}$ is proportional to $g_{R}$,
by tuning down $g_{R}$ one can obtain arbitrarily small $g_{{\rm eff}}$
to circumvent all constraints presented in Figs.~\ref{fig:bound}
and \ref{fig:bound-1}. On the other hand, if $g_{R}$ is very small,
then the lower bounds of $m_{Z'}$ discussed in Sec.~\ref{sec:mass} will also
be alleviated, implying that the dark photon could be very light.
Taking Eqs.~\eqref{eq:geffw2f}, \eqref{eq:geffz2f} and \eqref{eq:x-9},
we plot the blue lines in Figs.~\ref{fig:bound} and \ref{fig:bound-1}
with $m_{D}=v=246$ GeV and $g_{R}$ varying from 0 to $4\pi$. The
space below the blues lines is the theoretically favored region if
only the Dirac mass term breaks the $U(1)_{R}$ symmetry. This applies
to the UV complete model in Sec.~\ref{sub:modelB}. 

If the Majorana mass term also breaks the $U(1)_{R}$ symmetry, then
the lower bound of $m_{Z'}$ is set by Eq.~\eqref{eq:x-8} instead
of Eq.~\eqref{eq:x-9}. In the standard type I seesaw, we have $M_{R}\sim m_{D}^{2}/m_{\nu}$
which implies that for $m_{\nu}=0.1$ eV and $m_{D}=246$ GeV, the
$U(1)_{R}$ symmetry breaks at a high energy scale around $10^{14}$
GeV. For this case, we plot the green curve in Fig.~\ref{fig:bound}.
As shown in Fig.~\ref{fig:bound}, even though with $g_{R}\lesssim10^{-11}$
the mass of $m_{Z'}$ could be below the electroweak scale or lower,
the effective coupling is many orders of magnitude below any of known
experimental limits. 

The inaccessibly large $m_{Z'}$ of the green curve is due to the
underlying connection between $m_{\nu}$ and $M_{R}$ in the standard
type I seesaw. 
In some alternative neutrino mass models such as inverse seesaw~\cite{Mohapatra:1986bd}, the scale of $M_{R}$ is decoupled
from $m_{\nu}$, which 
allows for a sizable $\nu_{L}$-$\nu_{R}$
mixing even when $M_{R}$ is reduced to the TeV scale, and has motivated
many studies on collider searches for right-handed neutrinos---see Ref.~\cite{Deppisch:2015qwa} for a review.
 Here for illustration we simply set $M_{R}=m_{D}/\sin\theta$ with
$m_{D}=246$ GeV and $\sin\theta=10^{-2}$, which ensures that $\nu_{R}$
is sufficiently heavy  to avoid all current collider bounds. The possibility of collider-accessible $\nu_R$  involves more complicated phenomenology which is beyond the scope of this work.
The strength of $g_{{\rm eff}}$ and the lower bound of $m_{Z'}$
in this case is presented by the orange lines in Figs.~\ref{fig:bound}
and \ref{fig:bound-1}.

Now confronting  the theoretically favored $g_{{\rm eff}}$ and $m_{Z'}$
of the aforementioned three scenarios with the experimental limits,
we can see that only when the $U(1)_{R}$ breaking scale is determined
by $m_{D}$ or $M_{R}=m_{D}/\sin\theta$ with sizable $\sin\theta$,
the $\nu_{R}$-philic dark photon could be of phenomenological interest.
The former could potentially give rise to observable effects in long-range
force searches, astrophysical observations, beam dump  and collider
experiments. The latter, albeit beyond the current collider bounds,
might be of importance to future collider searches. In addition, the
SHiP experiment will be able to considerably dig into the parameter
space of the latter.

\section{Conclusion}\label{sec:concl}
The $\nu_R$-phillic dark photon $Z'$ which arises from a hidden $U(1)_R$ gauge symmetry and at the tree-level couples only to the right-handed neutrinos, interacts weakly with SM particles via loop-level processes---see Fig.~\ref{fig:WZ}.
Assuming the most general Dirac and Majorana mass matrices, we have derived loop-induced couplings of $Z'$ to charged leptons and quarks. The results are given in Eqs.~\eqref{eq:x-2} and \eqref{eq:x-5}, which are applied to a few examples including a UV complete model. 
For a special case with three $\nu_L$ and three $\nu_R$, the  loop-induced coupling are given by Eqs.\eqref{eq:geffw3f} and \eqref{eq:geffz3f}.
We have also discussed potential connections between the mass $m_Z'$ and the gauge coupling $g_R$ from the point of view of technical naturalness, which implies that $m_Z'$ should be generally above the lower bound in Eq.~\eqref{eq:x-8} if $M_{R}$ breaks $U(1)_R$, or the bound in Eq.~\eqref{eq:x-9} if only $m_D$ breaks the symmetry.

The theoretically favored values of the loop-induced couplings are confronted with experimental constraints and prospects in Figs.~\ref{fig:bound} and \ref{fig:bound-1}.
We find that the magnitude of loop-induced couplings allows current experiments to put noteworthy constraints on it. Future beam dump experiments like SHiP  and FASER together with upgraded collider searches will have substantially improved sensitivity on such a dark photon.

Hence as the answer to the question proposed in the title, we conclude
that the $\nu_{R}$-philic dark photon might not be inaccessibly dark
and could be of importance to a variety of experiments!

\begin{acknowledgments}
    \noindent
    The work of G.C. is supported in part by the US Department of Energy under Grant No. DE-SC0017987 and also in part by the McDonnell Center for the Space Sciences. X.J.X is supported by the ``Probing dark matter with neutrinos'' ULB-ARC convention and by the F.R.S./FNRS under the Excellence of Science (EoS) project No. 30820817 - be.h ``The $H$ boson gateway to physics beyond the Standard Model''. We acknowledge the use of {\tt Package-X} \cite{Patel:2015tea}, which is a great tool to simplify the loop calculations in this work.
\end{acknowledgments}

\appendix

\section{Explicit calculation of loop diagrams}\label{appendix:A}

In this appendix, we compute loop diagrams presented in Fig.~\ref{fig:WZ} in the mass basis.
In the main text, we use two-component Weyl spinors for conceptual simplicity. 
However, technically it is more convenient to convert them to four-component Dirac/Majorana spinors  so that the standard trace technology can be employed. Following the same convention as Ref.~\cite{Xu:2020qek}, we rewrite Eq.~(\ref{eq:m-11}) as
\begin{equation}
{\cal L}\supset (G_{Z})^{ij}Z_{\mu}\overline{\psi_{i}}\gamma_{L}^{\mu}\psi_{j}
+(G_{R})^{ij}Z'_{\mu}\overline{\psi_{i}}\gamma_{L}^{\mu}\psi_{j}
+\left[(G_{W})^{\alpha i}W_{\mu}^{-}\overline{\psi_{\alpha}}\gamma_{L}^{\mu}\psi_{i}
+{\rm h.c.}\right],\label{eq:m-20}
\end{equation}
where $P_L= \frac{1}{2}(1-\gamma_5)$,  $\gamma_{L}^{\mu}\equiv \gamma^{\mu}P_L$, and
\begin{equation}
\psi_{\alpha}=\left(\begin{array}{c}
    \ell_{L,\alpha}\\[2mm]
    \ell_{R,\alpha}^{\dagger}
    \end{array}\right),\ 
\psi_{i}\equiv\left(\begin{array}{c}
    \nu_{i}\\[2mm]
    \nu_{i}^{\dagger}
    \end{array}\right).    
\end{equation}
For simplicity,
we symbolically denote the relevant product of neutrino-gauge couplings by $G_X$ (it may stands for different quantities in different diagrams), which will be replaced by specific couplings when actually used.

\subsection{The $Z$ diagram}\label{sec:Z1}
The diagram is presented in the upper right panel in Fig.~\ref{fig:WZ}.
We first compute the vacuum polarization part of the diagram  (i.e. without the external fermion lines):
\begin{equation}
    i\mathcal{M}_{\mu\nu}=  G_X\int \frac{d^4k}{(2\pi)^4} \text{Tr}\left[ \ \gamma_\mu P_L \Delta_j(q-k) \   \gamma_\nu P_L \Delta_i(k)\right],\label{eq:circle}
\end{equation}
where $q$ is the momentum of $Z'$ and 
\begin{equation}
    \ \Delta_i(p)=\frac{i}{\slashed{p}-m_i}.
\end{equation}
Taking into account the Lorentz structure of the amplitude, this can be further decomposed as :
\begin{equation}
    i\mathcal{M}_{\mu\nu}= -\frac{iG_X}{16 \pi^2} \left[\mathcal{F}_1(m_i,m_j,q^2)\ {q_\mu q_\nu} + \mathcal{F}_2(m_i,m_j,q^2)\ g_{\mu\nu}\right],
\end{equation}
where 
\begin{align}
    \mathcal{F}_1(m_i,m_j,q^2)  \ = \ & \frac{5m_i^4-22m_i^2m_j^2+5m_j^4}{9(m_i^2-m_j^2)^2} + \frac{2m_j^4(3m_i^2-m_j^2) }{3(m_i^2-m_j^2)^3 } \log\left(\frac{m_i^2}{m_j^2}\right) \nonumber \\ &   + \ \frac{2}{3}\left[ \frac{1}{\epsilon}+\log\left(\frac{\mu^2}{m_i^2}\right)\right] + \mathcal{O}(q^2)\ , \label{eq:f1z}\\
    \mathcal{F}_2(m_i,m_j,q^2)  \ = \ & \frac{m_i^2+m_j^2}{2} - \frac{m_j^4}{(m_i^2-m_j^2)} \log\left(\frac{m_i^2}{m_j^2}\right) \nonumber \\ &   + \ {(m_i^2+m_j^2)\left[ \frac{1}{\epsilon}+\log\left(\frac{\mu^2}{m_i^2}\right)\right]} + \mathcal{O}(q^2)\ .\label{eq:f2z}
\end{align}
The full amplitude of the $Z$ diagram can be written as 
\begin{equation}
    i\mathcal{M}_{Z}= -i\ G_X \int \frac{d^4k}{(2\pi)^4} \text{Tr}\left[\ \gamma_\mu P_L \Delta_j(q-k) \   \gamma_\rho P_L \Delta_i(k)\right] \Delta_Z^{\rho\nu}(q) \ \overline{u(p_1)} \gamma_\nu P_{L/R} u(p_2),\label{eq:mz1}
\end{equation}
where the most general form  of $\Delta_Z^{\mu\nu}(q)$ in $R_\xi$ gauges is
\begin{equation}
    \Delta_Z^{\mu\nu}(q) = \frac{-i}{q^2-m_Z^2}\left[ g^{\mu\nu}- \frac{q^\mu q^\nu}{q^2-\xi m_Z^2}(1-\xi)\right].
    \label{eq:Zprop}
\end{equation}
We proceed with the unitarity gauge  corresponding to  $\xi\rightarrow\infty$, and the soft-scattering limit  $q\ll m_Z$:
\begin{equation}
    \Delta_Z^{\mu\nu}(k)  \xrightarrow{\xi\rightarrow\infty,\ q\ll m_Z}  \frac{ig^{\mu\nu}}{m_Z^2}.
    \label{eq:Zprop2}
\end{equation}
By applying the result of Eq.~\eqref{eq:circle} to Eq.~\eqref{eq:mz1}, we obtain
\begin{equation}
    i\mathcal{M}_{Z}=-i\ \frac{G_X}{16 \pi^2m_Z^2} \left[\mathcal{F}_1(m_i,m_j,q^2)\ q_\mu q_\nu + \mathcal{F}_2(m_i,m_j,q^2)\ g_{\mu\nu}\right] \overline{u(p_1)} \gamma^\nu P_{L/R} u(p_2),
\end{equation}
where $\mathcal{F}_1$ and $\mathcal{F}_2$  were already given in Eqs.~\eqref{eq:f1z} and \eqref{eq:f2z}, respectively.

\subsection{The $W$ diagram}\label{sec:W}
The diagram is presented in the upper left panel in Fig.~\ref{fig:WZ}.
The amplitude reads:
\begin{equation}
    i\mathcal{M}_{W}= -i\ G_X\int \frac{d^4k}{(2\pi)^4} \overline{u(p_1)}\gamma^\nu P_L \Delta_j(k-p_1)  \gamma^\rho P_L\Delta_i(p_2-k)  \gamma^\mu P_L u(p_2) \Delta^W_{\mu\nu}(k),
\end{equation}
where
\begin{equation}
    \Delta_i(p)=\frac{i}{\slashed{p}-m_i},
\end{equation}
\begin{equation}
    \Delta^W_{\mu\nu}(k) = \frac{-i}{k^2-m_W^2}\left[ g_{\mu\nu}- \frac{k_\mu k_\nu}{k^2-\xi m_W^2}(1-\xi)\right] .
    \label{eq:wprop}
\end{equation}
Similar to the $Z$ diagram, we take the unitarity gauge ($\xi\rightarrow \infty$) and the soft-scattering limit ($q\rightarrow 0$).
The quantity in the loop integral is proportional to 
\begin{equation}
    \int \frac{d^4k}{(2\pi)^4} \gamma^\nu P_L \Delta_j(k-p_1) \gamma^\rho P_L\Delta_i(p_2-k)  \gamma^\mu P_L \Delta^W_{\mu\nu}(k) \equiv C_a \gamma^\rho P_L + C_b P_L p_{1}^{\rho} + C_c P_L p_{2}^{\rho}.
\end{equation}
Here ($C_a,C_b,C_c$) are functions of scalar invariants $p_1^2$ and $p_2^2$. 
The last two terms are suppressed when imposing the on-shell conditions. 
Focusing only on the $\gamma^\rho P_L$ term, we obtain
\begin{equation}
    i\mathcal{M}_{W}= i \frac{G_X}{16\pi^2} \mathcal{F}(m_i,m_j) \ \overline{u(p_1)} \gamma^\rho P_L u(p_2),
\end{equation}
where
\begin{align}
    \mathcal{F}(m_i,m_j)  \ = \ & \frac{2m_i^2+2m_j^2+3m_W^2}{2 m_W^2} + \frac{m_j^4\log\left({m_j^2}/{m_W^2}\right)-m_i^4\log\left({m_i^2}/{m_W^2}\right)}{(m_i^2-m_j^2)\ m_W^2} \nonumber \\ &   + \ \frac{m_i^2+m_j^2}{m_W^2} \left[ \frac{1}{\epsilon}+\log\left(\frac{\mu^2}{m_W^2}\right)\right]\,.
\end{align}

\bibliographystyle{JHEP}
\bibliography{loop}

\end{document}